\input{aipcheck}

\documentclass[
    ,final            
  ]
  {aipproc}

\layoutstyle{6x9}

\begin{document}

\title{$B_s$, $B_c$ and $b$-baryons}


\author{Vaia Papadimitriou\\
(for the Tevatron, LEP and PEPII experimental collaborations)}{
  address={Fermi National Accelerator Laboratory, 
P.O. Box 500 Batavia, Illinois, 60510}
}



\begin{abstract}
 We present the latest measurements on masses, lifetimes and branching 
fractions for the $B_s$ and $B_c$ mesons as well as for $b$-baryons. For the
$B_s$ meson we discuss as well the latest results on mixing. 
These results were produced by the CDF and D0 experiments at Fermilab
or by earlier LEP and PEPII experiments.

\end{abstract}

\maketitle


\section{INTRODUCTION}

  In this paper we are reviewing recent results on the properties of heavy 
$B$ mesons and baryons. These include mass, lifetime and branching fraction 
measurements. For the $B_s$ meson we are also reviewing mixing.
The majority of the experimental results reported in this review come 
from the CDF and D0 experiments at Fermilab recording 
$p\bar{p}$ collisions at $\sqrt{s} = $1.96 TeV. Some of the reported results 
come as well from earlier $e^+e^-$ experiments at LEP and PEPII.

This paper is organized as follows. In section 1 we report briefly on the 
performance of the Tevatron, the main features of the CDF and D0 detectors as
well as their $B$ physics data sets.
In sections 2 and 3 we describe results 
on the $B_s$ and $B_c$ mesons respectively, and in section 4 we describe results
on $b$-baryons. In section 5 we discuss conclusions and prospects.

\section{1. Tevatron and the CDF and D0 data samples}

From August 1992 to February 1996 (Run I) the CDF and D0
detectors collected data samples of approximately 110 $pb^{-1}$ each
of $p\bar{p}$ collisions at $\sqrt{s} =$ 1.8 TeV. In Run I the crossing 
time was 3.5 $\mu s$ for 6 bunches and the typical luminosity of order
$10^{31}$ cm$^{-2}$sec$^{-1}$. 
Run II physics quality data started in March 2002. By September 2005, the
Tevatron has delivered more than 1 fb$^{-1}$ of data to each of the CDF 
and D0 detectors
 of $p\bar{p}$ collisions at $\sqrt{s} =$ 1.96 TeV. 
In Run II the crossing 
time is 396 ns for 36 bunches and the typical luminosity so far of order
$10^{32}$ cm$^{-2}$sec$^{-1}$.    

 For the $B$ physics results reported here, CDF uses up to 
360 $pb^{-1}$ of data while D0 uses up to 490 $pb^{-1}$ of data. 
These data sets are collected with three major trigger categories: 
dilepton triggers, single lepton triggers and triggers with displaced 
vertices. The dilepton triggers provide millions of $J/\psi$'s, $\sim$ 80\% 
of which 
come from $b$ decays. The single lepton triggers provide a large sample of 
$b$-hadrons decaying semileptonically, and sample sizes of up to $\sim$ 100 
thousand
lepton plus $D$ events are achieved. The triggers with displaced vertices 
provide a large number of $b$-hadrons decaying hadronically. From these 
samples several thousand of fully reconstructed hadronic $b$ decays are 
obtained. At present only CDF implements this type of trigger.

\section{2. The $B_s$ meson}

\subsection{Mass and lifetime}

Using 220 pb$^{-1}$ of Run II data CDF was able to present the most precise
individual measurements, to date, for the masses of $B^+$, $B^0$, $B_s^0$ and
$\Lambda^0_b$ \cite{CDFbmasses}.These measurements were made with 
exclusively reconstructed final states containing a 
$J/\psi \rightarrow \mu^+\mu^-$, and the systematic uncertainties achieved
for the $B$ meson masses were of the order of one third of an MeV.

The $B_s^0$ mass measurement was made by reconstructing 185$\pm$13 signal 
events in the decay channel
$B_s^0 \rightarrow J/\psi \phi$ where $\phi \rightarrow K^+K^-$, and the mass
was measured to be $m(B_s^0) =$ 5366.01 $\pm$ 0.73(stat) $\pm$ 0.33(syst) 
MeV/$c^2$. This measurement has better uncertainty than the current 
world average (WA).

The lifetime of $b$-hadrons is governed primarily by the decay of the 
$b$ quark, however contributions from the spectator quarks can be up to
$\sim$ 15\%. Presently these spectator effects are mostly calculated in the 
framework of the Heavy Quark Expansion Theory \cite{HQET}. Theory errors on 
the ratio of the $B_s^0$ and $B_d^0$ lifetimes is of the order of 1\%.

A summary of new $B_s^0$ lifetime measurements by CDF and D0 is
 presented in Table 1. 
Using 400 pb$^{-1}$ of Run II data taken with a single lepton trigger D0
performed the best available, to date, measurement of the $B_s^0$ lifetime 
(see Fig. 1(left) and Table 1). With 360 pb$^{-1}$ of Run II data and 
triggering on 
displaced vertices CDF has measured for the first time the lifetime of
$B$ mesons using fully reconstructed hadronic decays like 
$B \rightarrow D\pi$ or $B \rightarrow D3\pi$; the small systematic error
indicates a good control of the turn-on of the secondary vertex 
trigger efficiency (see Fig. 1(right) and Table 1). 
The winter 2005 HFAG WA for $\tau (B_s^0)$ was 
1.479$\pm$0.044 $ps$ for $B_s^0 \rightarrow D_s^+ X$ decays and
1.404$\pm$0.066 $ps$ for fully exclusive 
$B_s^0 \rightarrow J/\psi \phi$ decays \cite{HFAG_2005}.

\begin{table}
\begin{tabular}{lrrrr}
\hline
  & \tablehead{1}{r}{b}{CDF $J/\psi$ modes\\240 pb$^{-1}$}
  & \tablehead{1}{r}{b}{D0 $J/\psi$ modes\\220 pb$^{-1}$}
  & \tablehead{1}{r}{b}{CDF hadronic\\360 pb$^{-1}$}
  & \tablehead{1}{r}{b}{D0 semileptonic\\400 pb$^{-1}$}\\   
\hline
$\tau (B_s^0)$ ps & 1.369$\pm$0.100 
& 1.444$\pm$0.096 & 1.598$\pm$0.097$\pm$0.017     
& 1.420$\pm$0.043$\pm$0.057 \\
$\tau (B_s^0)/\tau (B_d^0)$ & 0.890$\pm$0.072 & 0.980$\pm$0.073 
&   &  \\
\hline
\end{tabular}
\caption{Summary of new $B_s^0$ lifetime results}
\label{tab:a}
\end{table}


\begin{figure}
  \includegraphics[height=.3\textheight]{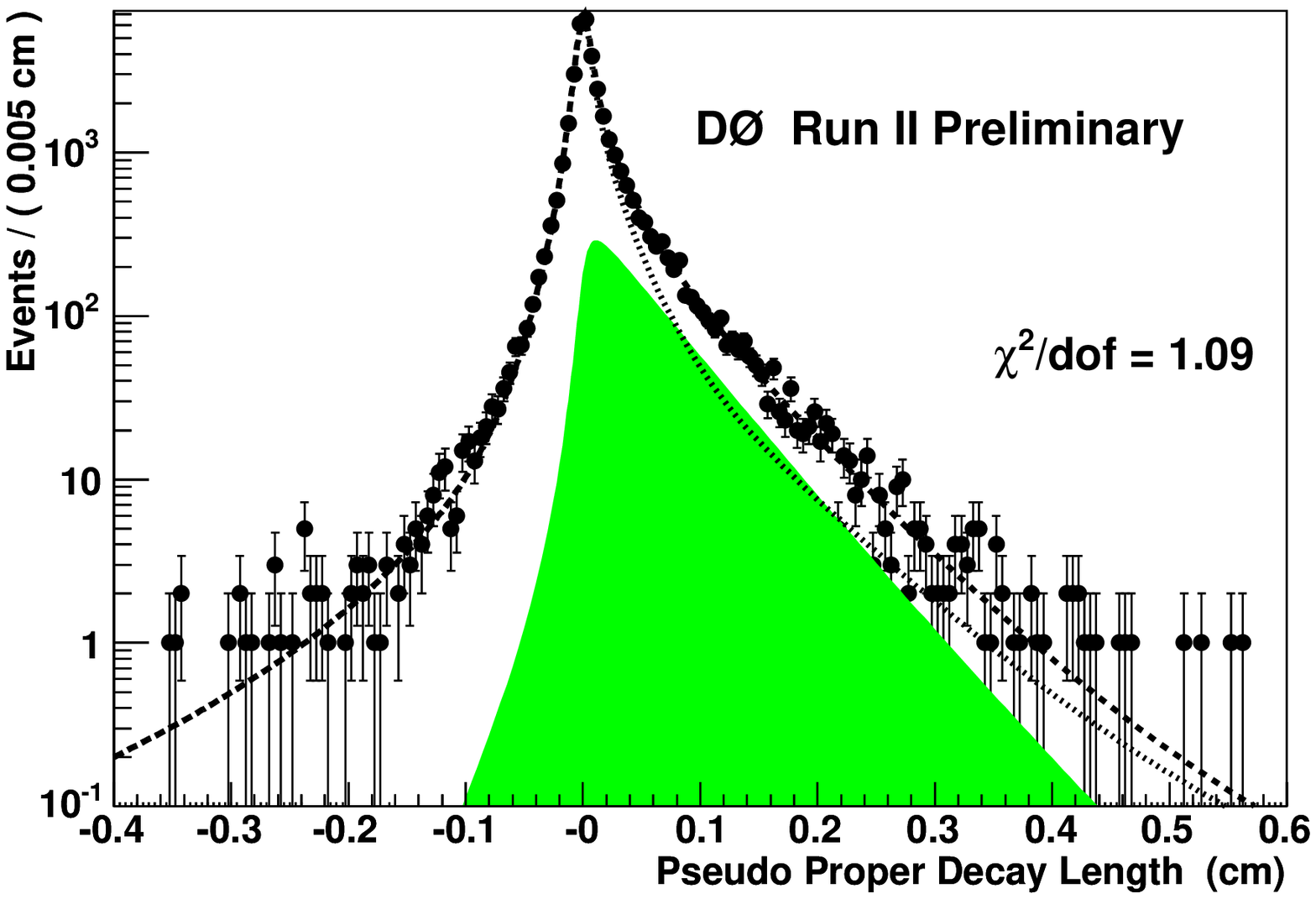}
 \hspace{-0.8cm}
 \includegraphics[height=.3\textheight]{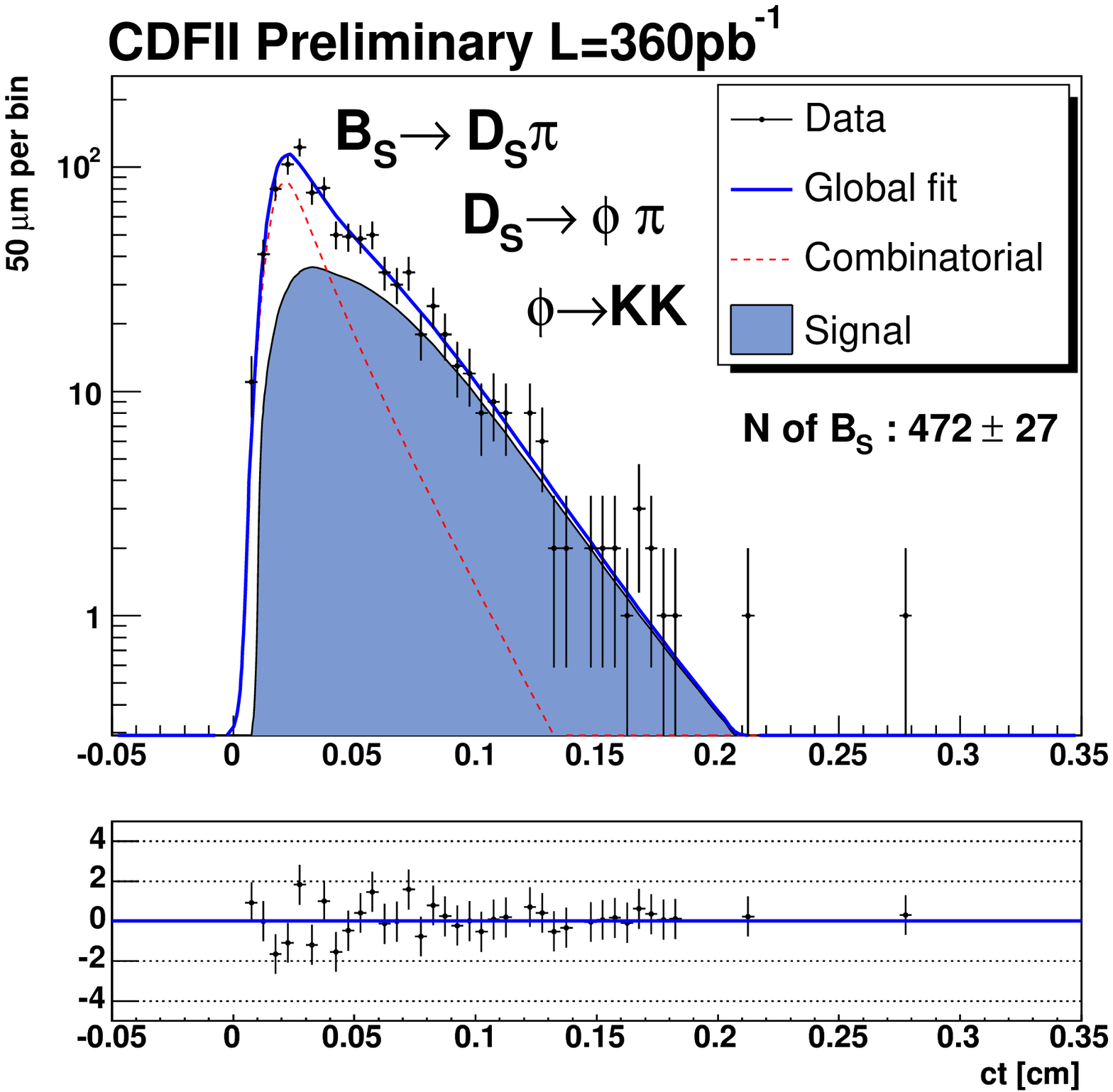}
  \caption{ (left) D0 pseudo-proper decay length distribution for $B_s^0$ 
semileptonic data with the result of the fit superimposed. The dotted curve
represents the combinatorial background and the filled histogram represents
the $B_s^0$ signal. (right) The lifetime projection from the combined 
mass-lifetime fit on the $B_s^0 \rightarrow D_s^-\pi^+$ decay mode at CDF.}
\end{figure}

\subsection{$B_s^0$ Mixing}

Within the framework of the standard model (SM), the $B_s^0$ mesons are 
supposed to mix 
in such a way that the mass and decay width differences between the heavy
and light eigenstates, $\Delta m_s \equiv M_H - M_L$ and $\Delta \Gamma _s \equiv 
\Gamma_L - \Gamma_H$, are sizeable.
The mixing phase $\delta \phi$ is small and to a good approximation 
the two mass eigenstates correspond to CP eigenstates. New phenomena may alter
$\delta \phi$, leading to a reduction of the observed 
$\Delta \Gamma_s /\bar{\Gamma_s}$ compared to the SM prediction.  
The frequency of the 
oscillation of the $B_s^0(B_d^0)$ flavor eigenstates into one another is 
proportional to the 
difference in mass between the two eigenstates and 
it is related to the CKM matrix element $|V_{ts}|(|V_{td}|)$.

Two different types of analyses have been performed so far to access 
$B_s^0$ mixing: 
the first one constrains  $\Delta m_s$ by measuring $\Delta \Gamma _s$
 while the second one is fitting the amplitude of
$B_s^0$ oscillations ($A(t) \sim D~ \rm x~ cos(\Delta m_s t)$). The status
of those two analyses will be discussed in the following subsections.

A measurement of both mixing frequencies $\Delta m_d$ and $\Delta m_s$ 
would yield a measurement of the ratio of the $|V_{td}|$ and $|V_{ts}|$ 
matrix elements, with a theoretical uncertainty of about 5\%, thus providing 
a strong constraint in global fits of the Unitarity triangle. If $\Delta m_s$
is too large to be directly measured, a measurement of $\Delta\Gamma_s$
could serve instead, along with $\Delta m_d$, in tests of the unitarity of 
the CKM matrix.

\subsubsection{$\Delta \Gamma_s$}

Within the standard model $\Delta m_s$ and $\Delta\Gamma_s$ are related
with the following theoretical relation \cite{Beneke}: 
\begin{equation}
\frac{\Delta m_s}{\Delta \Gamma_s} \approx \frac{2}{3\pi}\frac{m_t^2}{m_b^2}(1-
\frac{8}{3}\frac{m_c^2}{m_b^2})^{-1}h(\frac{m^2_t}{M_W^2}) \label{mix}
\end{equation}
So, according to \eqref{mix}, measuring $\Delta\Gamma_s$ leads to 
$\Delta m_s$ as well. 

In order to measure the decay width difference, $\Delta \Gamma_s$, we need
to disentagle the heavy and light $B_s^0$ mass eigenstates and measure
their lifetimes separately. Since the  $B_s^0$ mass eigenstates correspond 
also to 
the CP eigenstates, we can disentagle them by identifying the
CP even and CP odd contributions to the final states.
This is more staightforward when a 
pseudoscalar state decays to two vector mesons.
The decay $B_s^0 \rightarrow J/\psi \phi$ is a pseudoscalar to vector-vector
transition and is characterized by three amplitudes corresponding to
transitions in which the $J/\psi$ and $\phi$ have a relative orbital 
angular momentum $L$ of 0, 1, or 2. The observed final state particles for the
$B_s^0$, $\bar{B_s^0}$ decays ($\mu^+\mu^-K^+K^-$) have definite CP which 
depends on $L$.
Time-dependent angular analysis for 
$B_s \rightarrow J/\psi \phi$ together with lifetime measurements can help
separate heavy and light mass eigenstates.

\begin{table}
\begin{tabular}{lrrrr}
\hline
  & \tablehead{1}{r}{b}{$\Delta \Gamma_s/\bar{\Gamma_s}$}
  & \tablehead{1}{r}{b}{<$\tau$> ps}
  & \tablehead{1}{r}{b}{$\tau_L$ ps}
  & \tablehead{1}{r}{b}{$\tau_H$ ps}   \\
\hline
CDF & 0.65$^{+0.25}_{-0.33}$ & 1.40$^{+0.15}_{-0.13}$ & 1.05$^{+0.16}_{-0.13}$    & 2.07$^{+0.58}_{-0.46}$\\
D0 & 0.21$^{+0.33}_{-0.45}$ & 1.39$^{+0.15}_{-0.16}$  & 1.23$^{+0.16}_{-0.13}$   &  1.52$^{+0.39}_{-0.43}$ \\
\hline
\end{tabular}
\caption{Lifetimes and decay width difference for the heavy and light $B_s^0$
 states}
\label{tab:b}
\end{table}

In Table 2 we present the relative width difference between the heavy and light
mass eigenstates, $\Delta\Gamma_s/\bar{\Gamma_s}$, the average lifetime of the
($B_s^0$, $\bar{B_s^0}$) system and the mean lifetimes of the light and heavy
$B_s^0$ eigenstates as measured by the CDF \cite{CDFang} and D0 \cite{D0ang} 
experiments.  

\begin{figure}
  \includegraphics[height=.3\textheight]{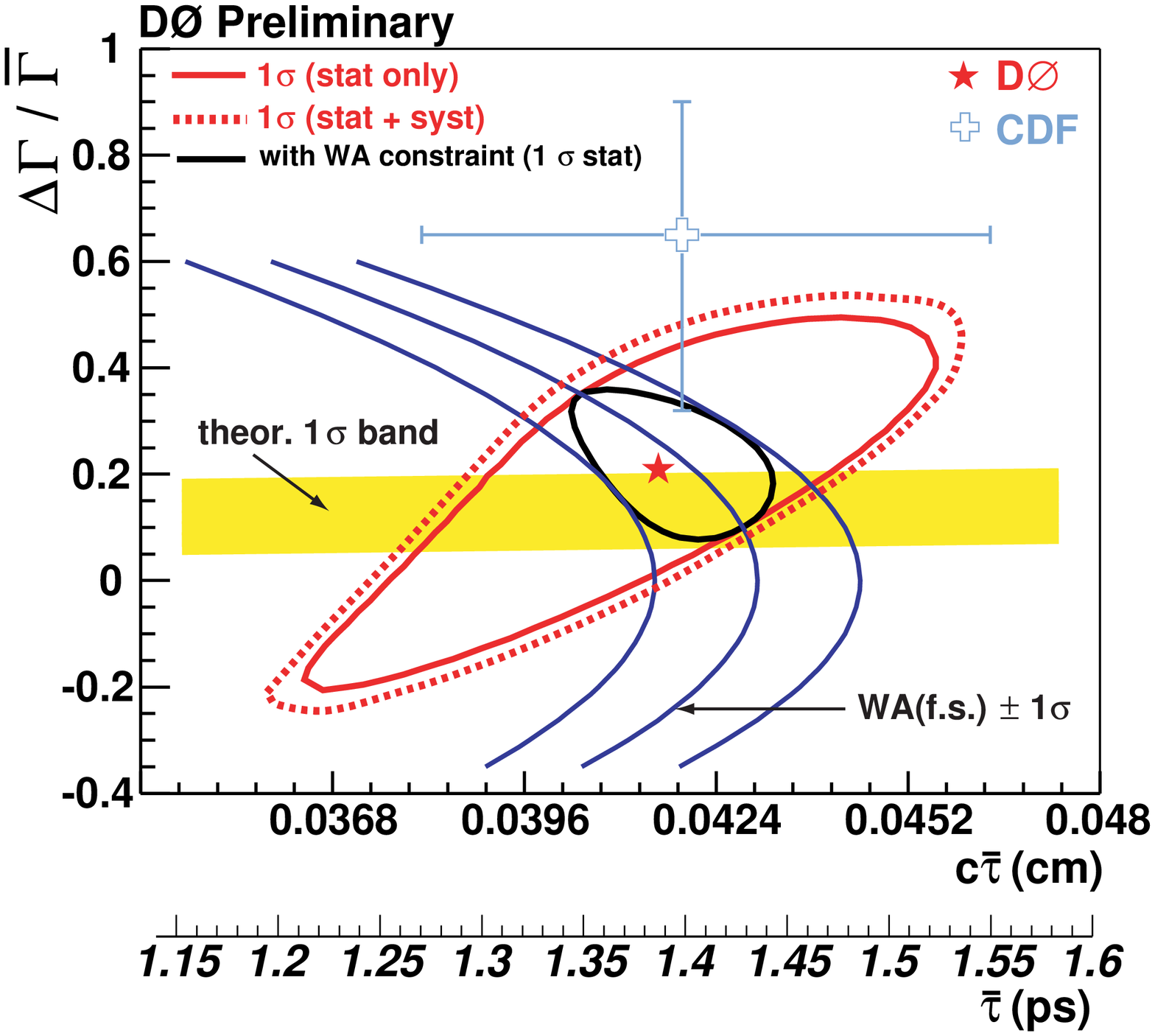}
\hspace{-0.1cm}
 \includegraphics[height=.28\textheight]{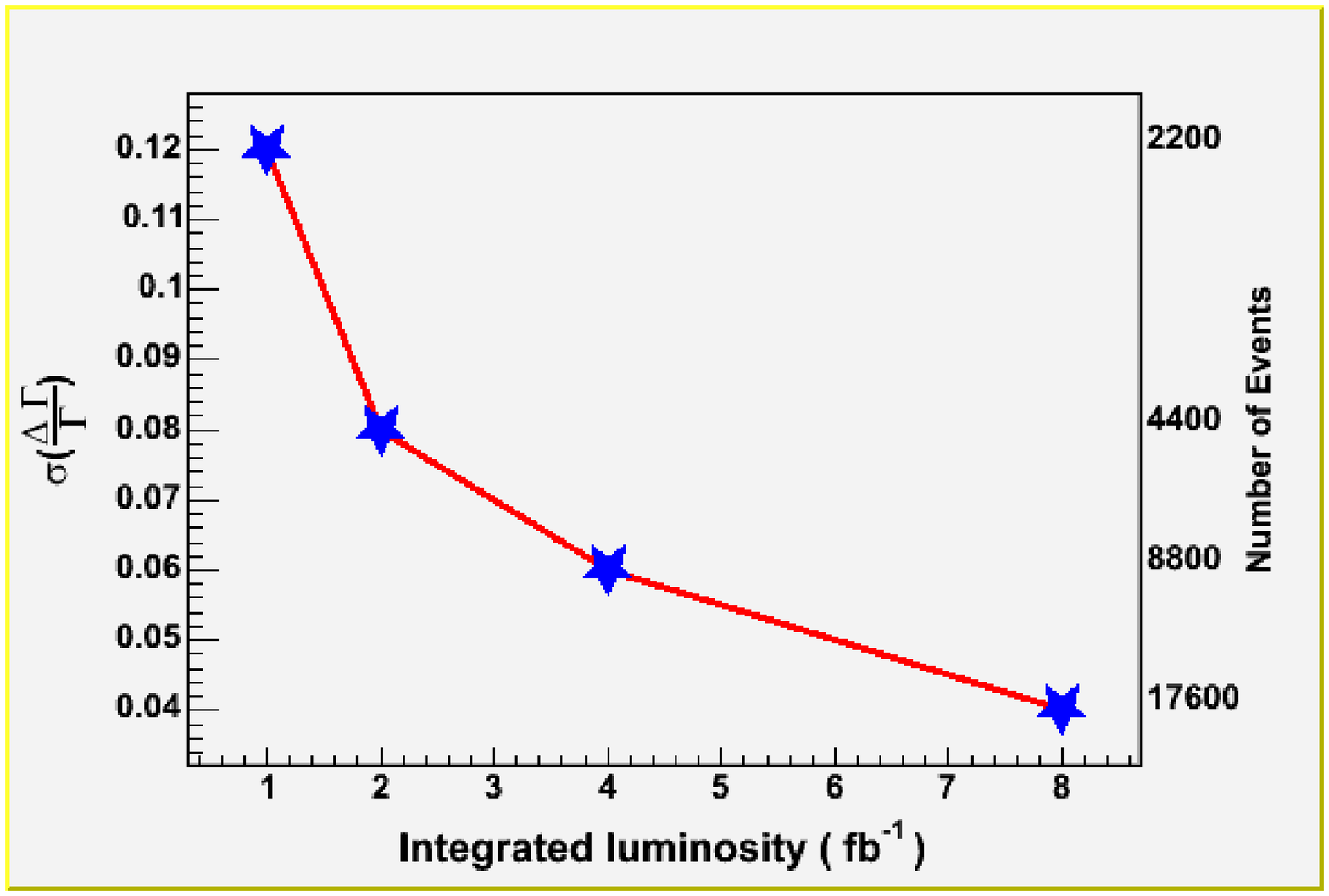}
  \caption{(left) $\Delta\Gamma_s/\bar{\Gamma_s}$ vs average lifetime. The D0
one-$\sigma$ contours are compared to a one-$\sigma$ band for the world average
 measurement based on flavor-specific decays. The CDF measurment as well 
as the SM theoretical prediction are also shown. 
(right) Projection of the uncertainty on $\Delta\Gamma_s/\bar{\Gamma_s}$ as a 
function of 
integrated luminosity per Tevatron experiment in Run II.}
\end{figure}

In Fig. 2(left) we show  $\Delta\Gamma_s/\bar{\Gamma_s}$ vs average lifetime
as measured by the CDF and D0 experiments and in comparison with the
theoretical expectation \cite{Dunietz}.
In Fig. 2(right) we show the expected precision on the measurement of 
$\Delta\Gamma_s/\bar{\Gamma_s}$ as a function of integrated luminosity
per Tevatron experiment. The estimate presented here is based on D0 
studies and the 
assumption that the CDF net sensitivity is about the same (CDF has a 
better mass resolution, hence less background under the signal, while D0 has 
a larger polar angle coverage) \cite{projection}.

\subsubsection{$\Delta m_s$}
 
Very precise measurements of $\Delta m_d$ have been available for some time,
and are currently dominated by the results of the $B$ 
factories \cite{factories}. Recent CDF and D0 measurements are consistent with
those results \cite{Bd_CDF_D0}. 

$\Delta m_s$ is constrained to be larger than 14.5 $ps^{-1}$ at 95\% CL from
previous work by the LEP experiments, SLD and CDF (Run I). On the basis of 
the same work   
the sensitivity (expected limit) is equal to 18.2 $ps^{-1}$. 
Both CDF \cite{CDF_Bsmix} and D0 \cite{D0_Bsmix} have recently obtained 
new limits on $\Delta m_s$.

Using 355 pb$^{-1}$ of Run II data CDF made two parallel measurements 
for $\Delta m_s$ using both fully 
reconstructed  
hadronic decays and semileptonic decays with a fully reconstructed $D_s^{\pm}$
meson. In the former case about 900 $B_s^0 \rightarrow D_s^{\pm}\pi^{\mp}$
were reconstructed after summing over three possible $D_s^{\pm}$
decay modes: $\phi \pi^{\pm}$, $K^{*0}K^{\pm}$ and $\pi^+\pi^-\pi^+$.
The semileptonic analysis yielded about 7500 
$B_s^0 \rightarrow D_s^{\pm}l^{\mp}\nu$ events summing over the three 
above mentioned 
$D_s^{\pm}$ decay channels, however the proper time resolution was worse
because of the incomplete knowledge of the decay kinematics due to the 
missing neutrino.
For the results presented here flavor tagging was performed using only opposite
-side taggers. The tag sign is provided either by the sign of the electron 
or muon or by the average weighed charge of the tracks in a jet.
A combined tagging power ($\epsilon D^2$) of the order of 1.4\%
was obtained.

CDF performed an amplitude scan as a function of $\Delta m_s$ \cite{Rouss} on 
both samples and then combined
them to obtain a 95\% CL limit of 7.9 $ps^{-1}$ and a sensitivity of 
8.4 $ps^{-1}$. In Fig. 3(left) we show the amplitude scan result which is 
clearly dominated by the statistical error.

Using 460 pb$^{-1}$ of Run II data D0 measured  $\Delta m_s$ in a large
semileptonic sample of $B_s^0 \rightarrow \mu^+ D_s^- X$ decays where the 
$D_s^-$ is fully reconstructed. 
The opposite-side muon tagging method was used for the initial
state flavor determination and the tagging power was of the order of 1.1\%.

In Fig. 3(right) we show the D0 amplitude scan as a function of $\Delta m_s$.
A 95\% CL limit of 5.0 $ps^{-1}$ and a sensitivity of 4.6 $ps^{-1}$ were 
obtained.

These new results have not improved yet the $\Delta m_s >$ 14.5 $ps^{-1}$
WA limit, but have extented the sensitivity to 18.5 $ps^{-1}$.  

In Fig. 4(left) we show the 5$\sigma$ observation projected sensitivity of 
$\Delta m_s$ for the combined hadronic and semileptonic CDF analyses
as a 
function of integrated luminosity per Tevatron experiment in 
Run II \cite{projection}. 
The improvements assumed are listed in the figure.
Similar projected sensitivities were assumed for the CDF and D0 experiments. 
The horizontal axis is interpreted as integrated luminosity per 
experiment under the assumption that only half of the delivered luminosity is
effectively usable by the experiments.


\begin{figure}
  \includegraphics[height=.3\textheight]{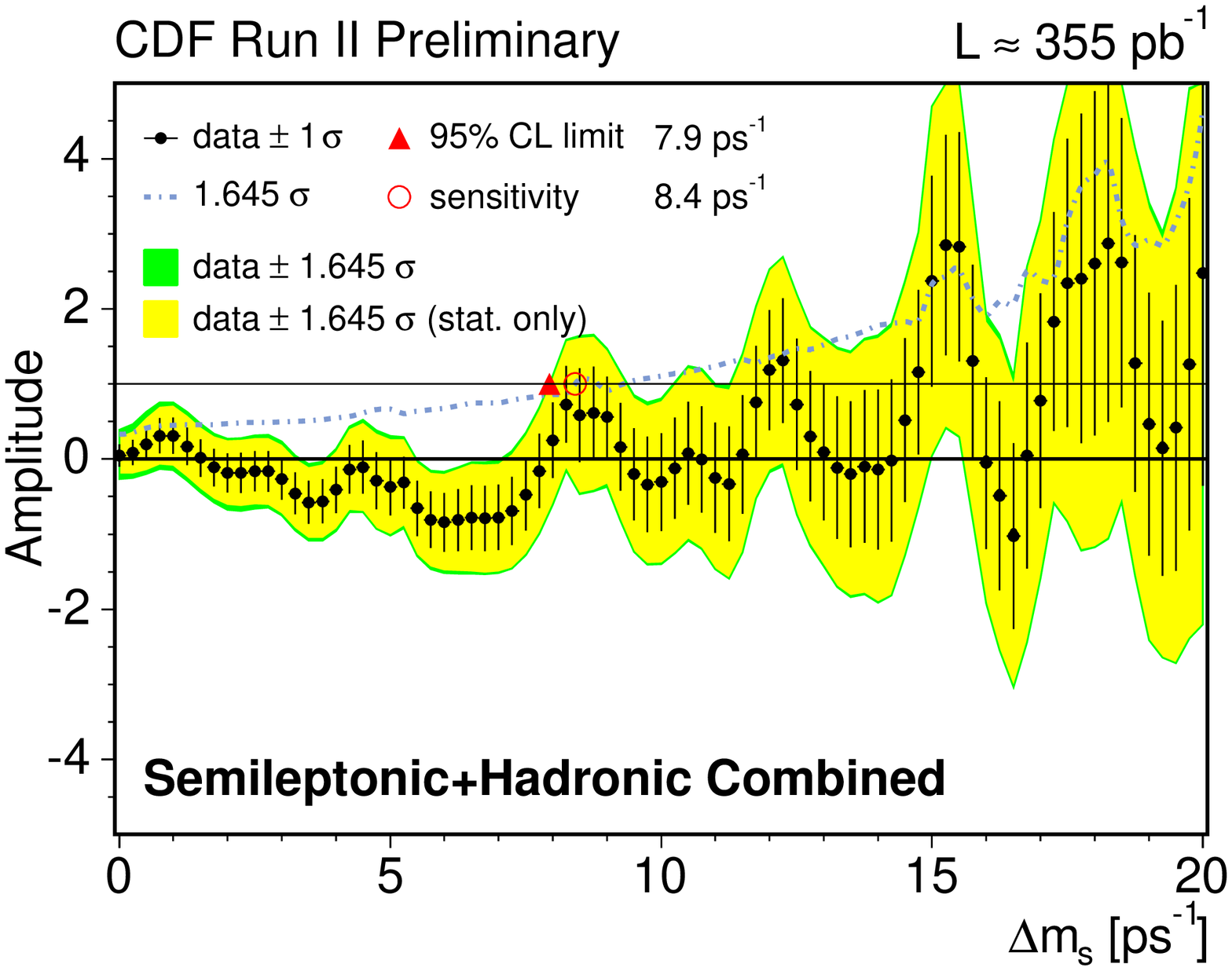}
 \hspace{-0.1cm}
 \includegraphics[height=.29\textheight]{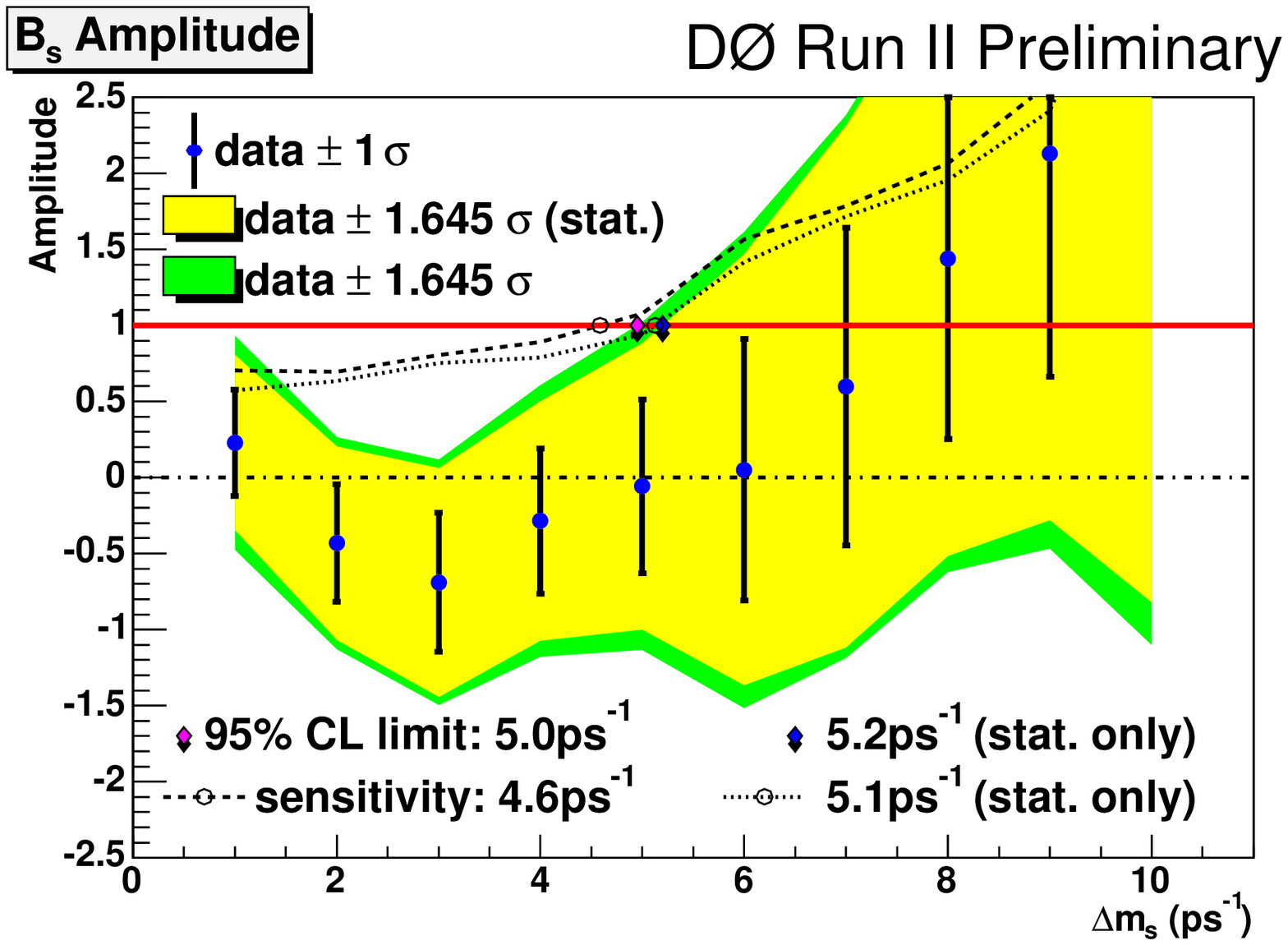}
  \caption{ (left) CDF amplitude scan combining hadronic and 
semileptonic $B_s^0$ signals. (right) D0 amplitude scan using semileptonic $B_s^0$ signals. }
\end{figure}

\begin{figure}
  \includegraphics[height=.3\textheight]{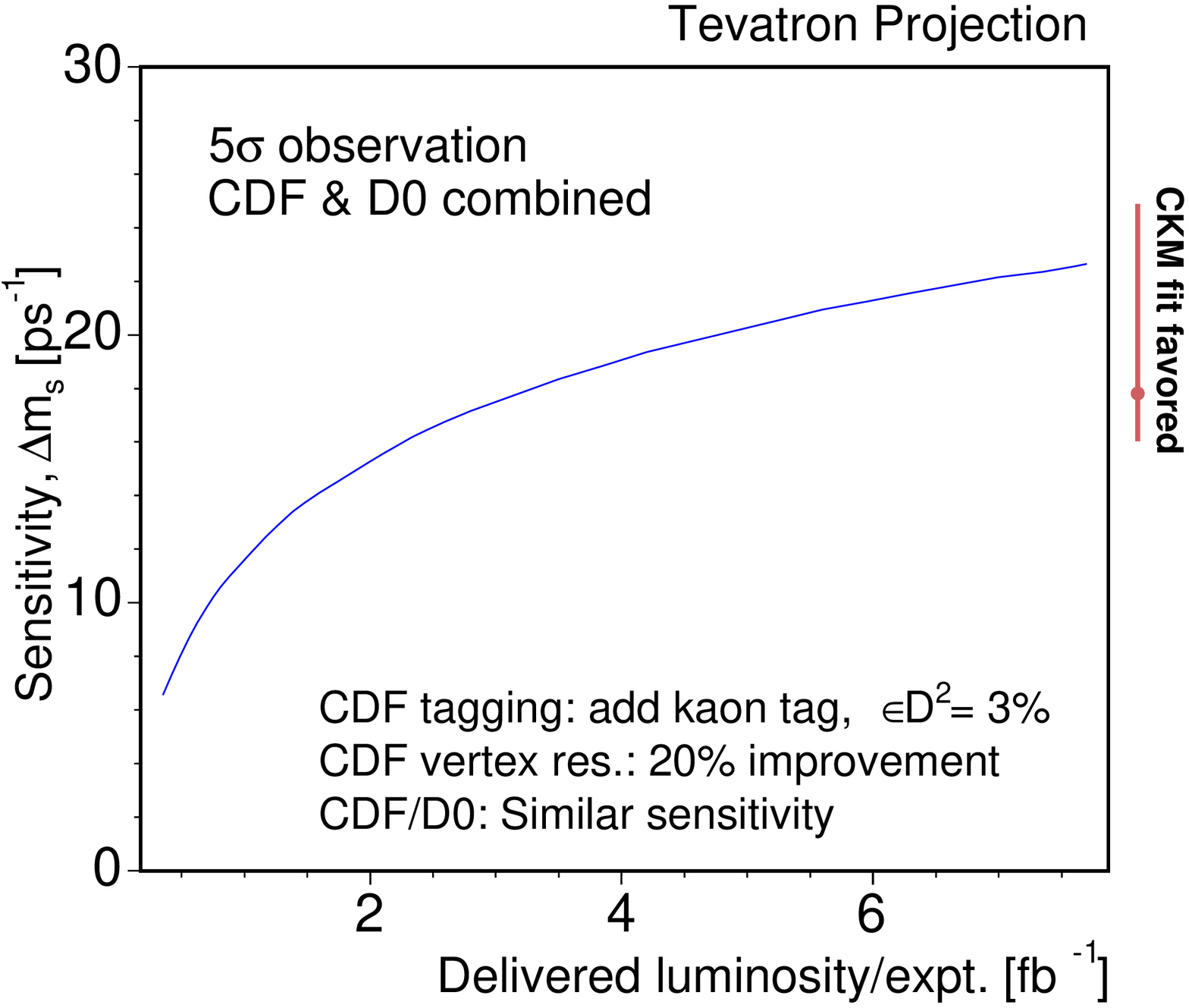}
 \includegraphics[height=.3\textheight]{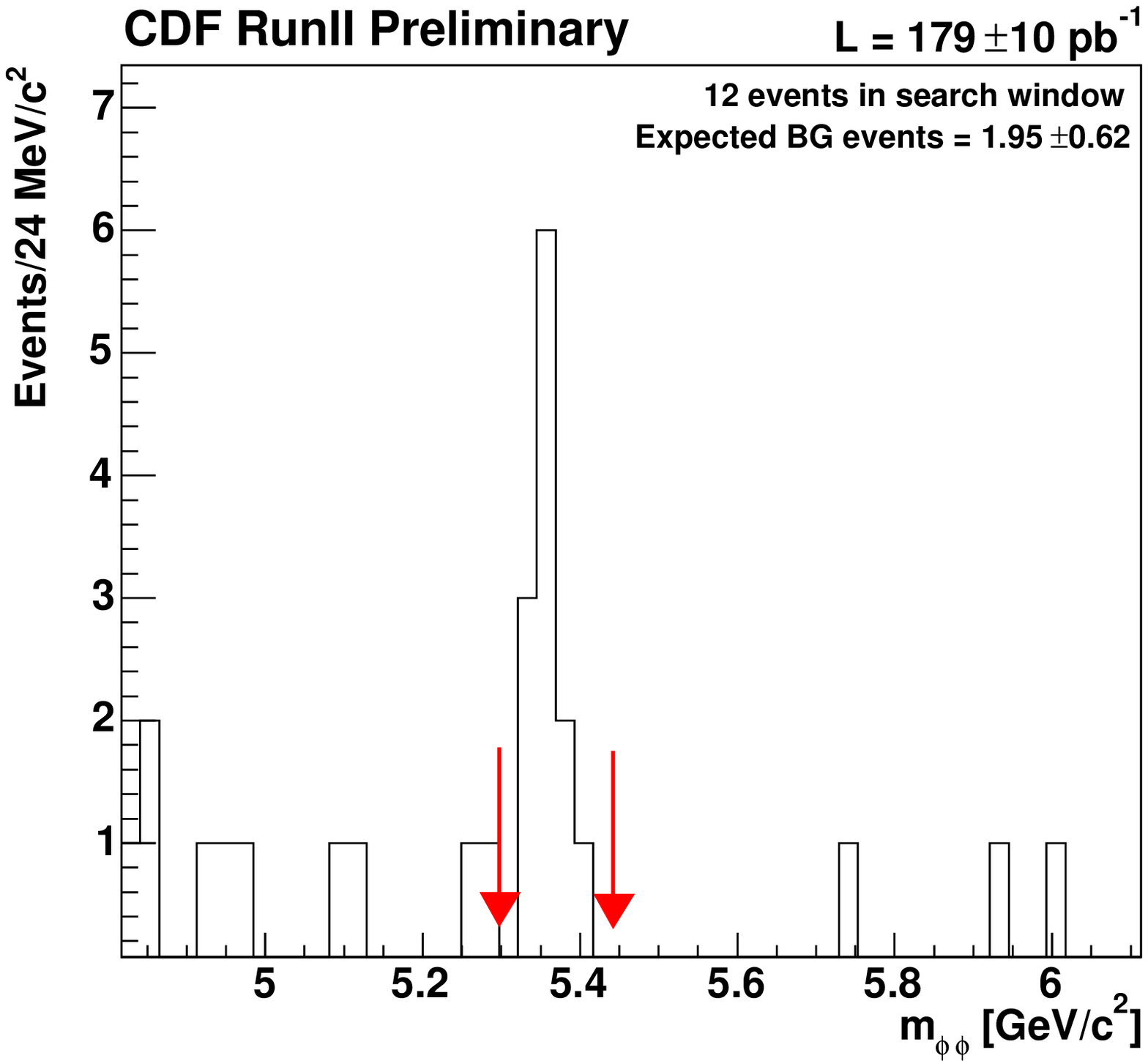}
  \caption{(left) $\Delta m_s$ sensitivity as a function of delivered 
luminosity per Tevatron experiment in Run II. (right) The CDF
K$^+$K$^-$K$^+$K$^-$ invariant mass distribution from the decay 
$B_s^0 \rightarrow \phi\phi$.}
\end{figure}

The value of $\Delta m_s$ predicted by SM on the basis of the analysis of all 
available $B$ and Kaon physics data is 
$\Delta m_s =$ 18.5$\pm$1.6 $ps^{-1}$, and the expected
95\% CL range 15.7 to 23 $ps^{-1}$ \cite{UTfit}.
All SM extensions predict a larger value, of at least 30 $ps^{-1}$. 

\subsection{Branching fractions - Rare decays}

CDF presented the first evidence of charmless decays of the $B_s^0$ meson,
$B_s^0 \rightarrow \phi\phi$ on the basis of 179 pb$^{-1}$ of 
Run II data. A ``blind'' search was performed fixing the selection requirements
and evaluating the combinatorial background from independent samples before 
examining the signal region in the data. The $B_s^0 \rightarrow \phi\phi$ 
candidate mass distribution is shown in Fig. 4(right). In a region of $\pm$72
MeV/c$^2$ around the world average $B_s^0$ mass, corresponding to a window
three times the expected mass resolution, 8 events were observed.
Using the $B_s^0 \rightarrow J/\psi \phi$ sample for normalization CDF derived 
 $Br (B_s^0 \rightarrow \phi\phi) = (14 ^{+6}_{-5} (stat) 
\pm 6 (syst)) \rm x 10 ^{-6}$. The corresponding theoretical predictions vary
in the range 18-37 x 10$^{-6}$. 

Using 360 pb$^{-1}$ of Run II data CDF presented as well the first 
observation of the decay 
$B_s^0 \rightarrow \psi \rm(2S)\phi$ in both the 
$\psi(2S) \rightarrow \mu^+\mu^-$ (5.02 $\sigma$) and 
$\psi(2S) \rightarrow J/\psi\pi^+\pi^-$ (4.16 $\sigma$) decays for a combined
significance of 6.35 $\sigma$.
Figure 5 shows the $\mu^+\mu^- K^+K^-$ invariant mass distribution with a fit
to the data superimposed. Using the $B_s^0 \rightarrow J/\psi \phi; 
J/\psi \rightarrow \mu^+\mu^-$ signal
as a control sample and for normalization, CDF derived
$Br (B_s^0 \rightarrow \psi(2S)\phi)/
Br (B_s^0 \rightarrow J/\psi\phi) =$ 0.52 $\pm$ 0.13 (stat) 
$\pm$ 0.06 (BR) $\pm$ 0.04 (syst).

Within SM, Flavor Changing Neutral Current (FCNC) decays are
highly suppressed and can only occur through higher 
order diagrams. The SM expectations for the branching ratios of 
$B_s^0 \rightarrow \mu^+\mu^-$ and $B_d^0 \rightarrow \mu^+\mu^-$
are $Br (B_s^0 \rightarrow \mu^+\mu^-) = (3.42 \pm 0.54) \rm x 10 ^{-9}$ and
$Br (B_d^0 \rightarrow \mu^+\mu^-) = (1.00 \pm 0.14) \rm x 10 ^{-10}$ \cite{Buch},
which are about two orders of magnitude smaller than the current experimental
sensitivity. However, new physics contributions can significantly enhance 
these branching fractions.

\begin{figure}
 \includegraphics[height=.3\textheight]{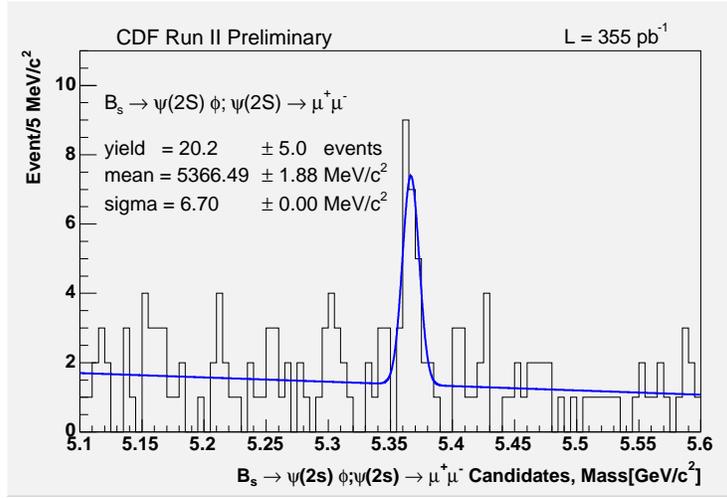}
  \caption{
Invariant mass distribution of $\mu^+\mu^- K^+K^-$ from the decay
$B_s^0 \rightarrow \psi(\rm 2S)\phi$. The fitting function is a 
single Gaussian for the signal and a first order polynomial for the background.}
\end{figure}

For the $ B_s^0 \rightarrow \mu^+\mu^-$, both CDF and D0 performed  ``blind'' 
search analyses and used the $B^{\pm} \rightarrow J/\psi K^{\pm}$
decay as a control sample and for normalization.
Using 240 pb$^{-1}$ of Run II data D0 found 4 $B_s^0$ candidates on an 
expected background of
4.3 $\pm$ 1.2 events and established an upper limit of
$Br (B_s^0 \rightarrow \mu^+\mu^-) < 3.7 \rm x 10 ^{-7}$ at 95\% CL.
Using 364 pb$^{-1}$ of Run II data CDF also searched for the 
$B_{s(d)}^0 \rightarrow \mu^+\mu^-$ decays.
As shown in Fig. 6(left) CDF observes no candidate events within the signal 
window
of $\pm$ 60 MeV ($\pm$2.5 $\sigma$) about the world average of $B_s^0$ or 
$B_d^0$ mass. They expected 0.81$\pm$0.12 and 0.66$\pm$0.13 events for the
central-central and central-extension channels respectively.
 They established limits of
$Br (B_s^0 \rightarrow \mu^+\mu^-) < 2.0 \rm x 10 ^{-7}$ 
and $Br (B_d^0 \rightarrow \mu^+\mu^-) < 4.9 \rm x 10 ^{-8}$ at 95\% CL.

Using a Bayesian integration technique and taking into account correlated and
uncorrelated systematic uncertainties the combined CDF and D0 limits are
$Br (B_s^0 \rightarrow \mu^+\mu^-) < 1.6 \rm x 10 ^{-7}$ and
$Br (B_d^0 \rightarrow \mu^+\mu^-) < 3.8 \rm x 10 ^{-8}$ at 95\%.
In Fig. 6(right) we show the expected $B_s^0 \rightarrow \mu^+\mu^-$ limit 
projection from CDF as a function of the delivered luminosity per Tevatron 
experiment in Run II.
The projection is based on the current CDF optimization of the selection 
requirements. A re-optimization at about 1 or 2 fb$^{-1}$
of luminosity could help improve this projection.

\begin{figure}
  \includegraphics[height=.3\textheight]{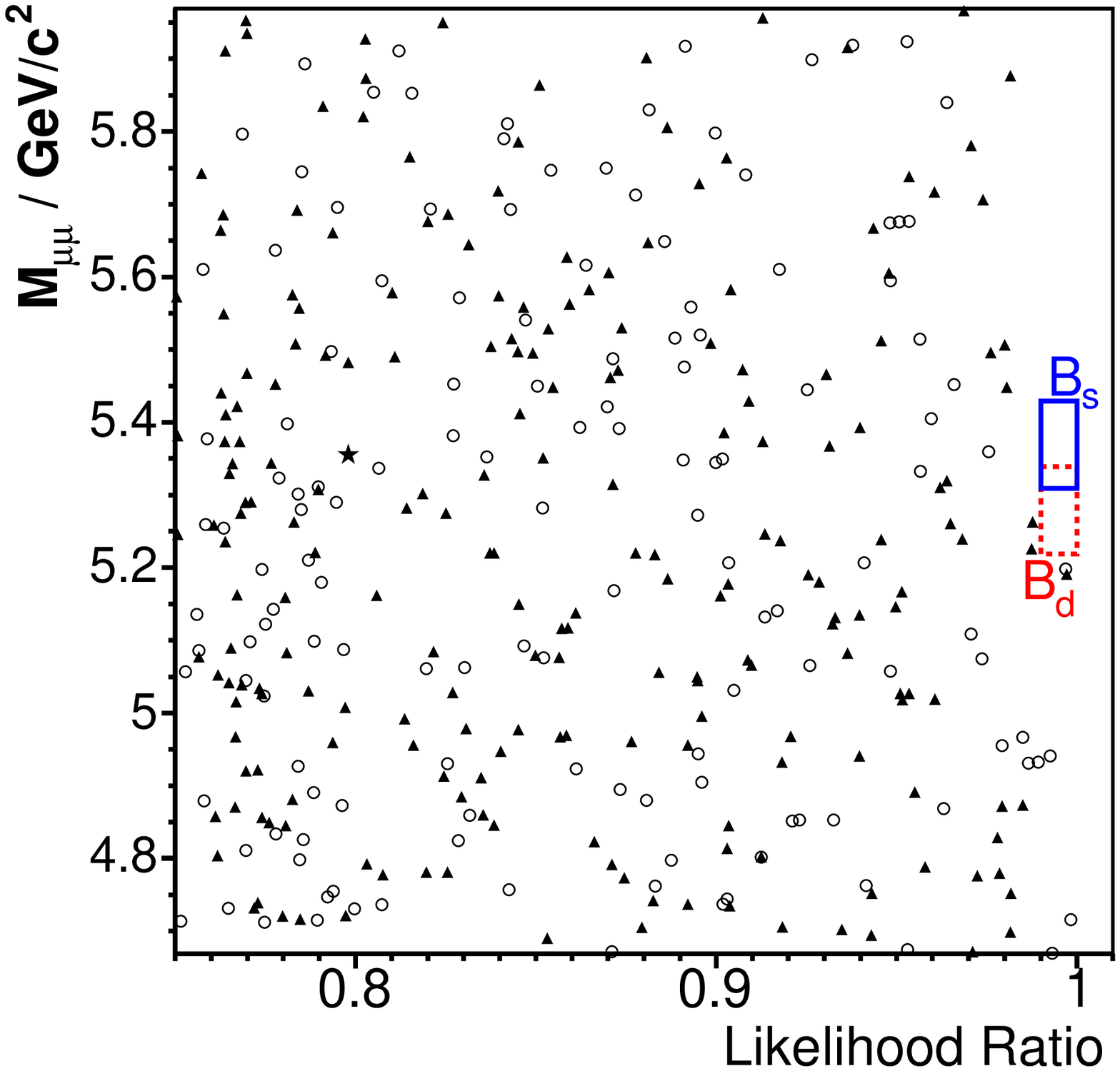}
\hspace{-0.3cm}
 \includegraphics[height=.31\textheight]{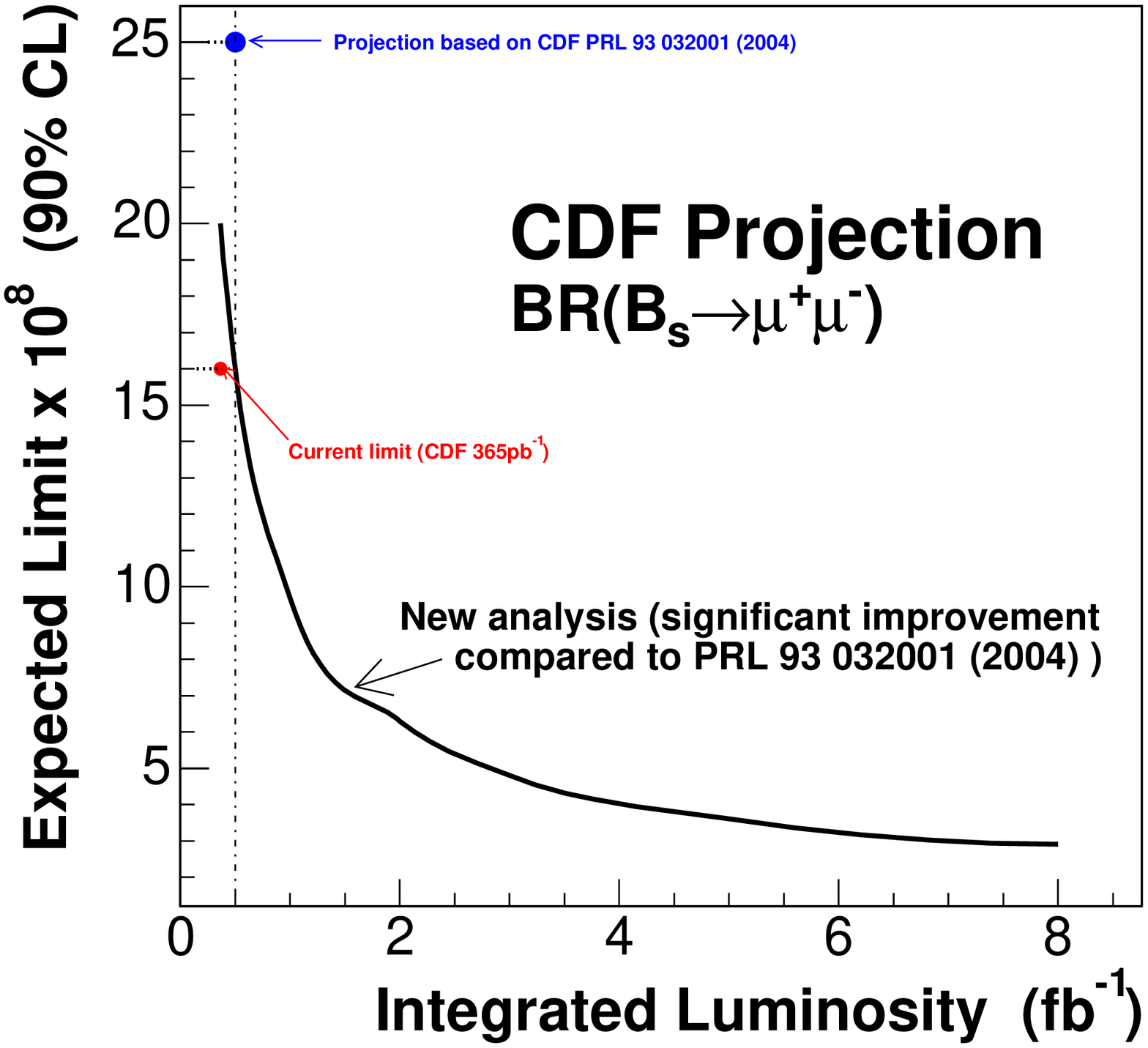}
  \caption{(left) The $\mu^+\mu^-$ invariant mass distribution versus 
likelihood ratio for events satisfying baseline requirements for 
central-central (solid triangle) and central-extension (open circle) channels. 
The $B_s^0$ (solid box) and $B_d^0$ 
(dashed box) signal regions are also shown. 
(right) The expected limit on the branching ratio of 
$B_s^0 \rightarrow \mu^+\mu^-$ as a function of delivered luminosity per
Tevatron experiment.}
\end{figure}

Using about 300 pb$^{-1}$ of Run II data D0 performed a ``blind'' search
for the decay $B_s^0 \rightarrow \mu^+\mu^- \phi$.
They expected 1.6$\pm$0.4 background events and they observed none in the 
search window.
They established a limit of    
$Br (B_s^0 \rightarrow \mu^+\mu^-\phi) < 4.1 \rm x 10 ^{-6}$ at 90\% CL.

\section{3. The $B_c$ meson}

The $B_c^{\pm}$ meson is made of a bottom-charm antiquark-quark pair.
 Nonrelativistic potential models predict the $\bar{b}$ and $c$ quarks
to be tightly bound with a ground state mass in the approximate
range 6200-6300 MeV/$c^2$ \cite{Rosner_91, Quigg_94,
Godfrey_04}. Recent QCD-based perturbative computations up 
to  ${\cal
O}(\alpha_s^4)$ 
predict $M(B_c)$ to be 6307 $\pm$ 17 MeV/$c^2$
\cite{Brambilla_2002}. Most recently,
a three-flavor lattice QCD calculation 
obtains 
$M(B_c) =$ 6304 $\pm$ 12(stat $\oplus$ syst) $^{+18}_{-0}$(cutoff effects)~MeV/$c^2$ \cite{lattice_2004}. 
The predicted $B_c^{\pm}$ lifetime is in the range 
of 0.4-1.4 ps \cite{bclife}. 

\subsection{Mass and lifetime}

 The CDF
collaboration made the first observation of the $B_c^{\pm}$ meson in the
semileptonic decay channels $B_c^{\pm} \rightarrow J/\psi l^{\pm} \nu_l X$,
in a sample of 110 pb$^{-1}$ of data in Run I \cite{CDFRunI_bc}. 
With a signal of
20.4$^{+6.2}_{-5.5}$ events, the $B_c^{\pm}$ mass and lifetime were measured 
to be 6.40 $\pm$ 0.39(stat) $\pm$ 0.13(syst) GeV/$c^2$ and 
0.46 $^{+0.18}_{-0.16}$ (stat) $\pm$ 0.03 (syst) ps respectively. In August 
2004, the D0
Collaboration reported a preliminary observation of a $B_c^{\pm}$ signal in
the decay channel $B_c^{\pm} \rightarrow J/\psi \mu^{\pm} \nu_{\mu} X$ in 
a sample
of 210 pb$^{-1}$ of Run II data \cite{D0RunII_bc}. On the basis of 
95 $\pm$ 12 $\pm$ 11 signal events
their analysis yielded a $B_c$ mass of  
5.95$^{+0.14}_{-0.13}$(stat) $\pm$ 0.34(syst) GeV/$c^2$ and a lifetime of
0.448 $^{+0.123}_{-0.096}$ (stat) $\pm$ 0.121 (syst) ps.

Using approximately 360 pb$^{-1}$ of Run II data CDF searched for the exclusive
decay mode $B_c^{\pm} \rightarrow J/\psi \pi^{\pm}$. 
The search window was chosen to correspond to the $\pm$2 standard deviation
region around the CDF Run I measurement of the $B_c^{\pm}$ mass; it was  
5.6-7.2 GeV/c$^2$, approximately 100 times wider than the expected 
$B_c^{\pm}$ mass resolution. A global unbinned likelihood fit to the 
$J/\psi \pi^{\pm}$ spectrum over the entire mass range yielded 
a $B_c^{\pm}$ mass of  
6285.7 $\pm$ 5.3(stat) $\pm$ 1.2(syst) MeV/$c^2$ for a 
signal of 14.6 $\pm$ 4.6 signal events on a
background of 7.1 $\pm$ 0.9 events within a region of $\pm$2 standard 
deviations from this mass value (see Fig. 7) \cite{CDFbcexc}.
 This peak is consistent with a narrow particle state which
decays weakly, and is interpreted as the first evidence for fully
reconstructed decays of the $B_c^{\pm}$ meson. 
The probability that a random background fluctuation would generate 
such a peak anywhere in the search window was calculated to be 0.012\%. 
The mass value agrees with
the much less precise mass values found in $B_c^{\pm}$ semileptonic decays.
There is also good agreement with recent theoretical predictions for
the $B_c^{\pm}$ mass 
around 6300 MeV/$c^2$~\cite{Brambilla_2002, lattice_2004}.

\begin{figure}
 \includegraphics[height=.3\textheight]{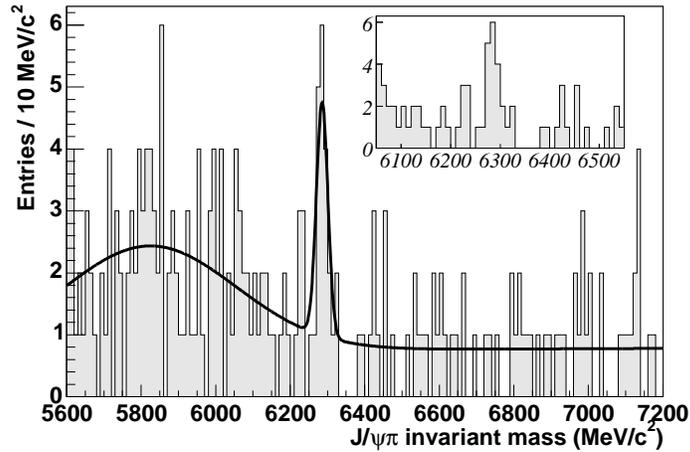}
  \caption{
The invariant mass distribution of the  
$J/\psi \pi^{\pm}$ candidates and results
of an unbinned likelihood fit in the search window. The inset shows the 
peak section of the distribution. The broad enhancement below 6.2 GeV/c$^2$
is attributable to partially reconstructed $B_c^{\pm}$ mesons.}
\end{figure}

\subsection{Branching fractions}

\begin{figure}
  \includegraphics[height=.28\textheight]{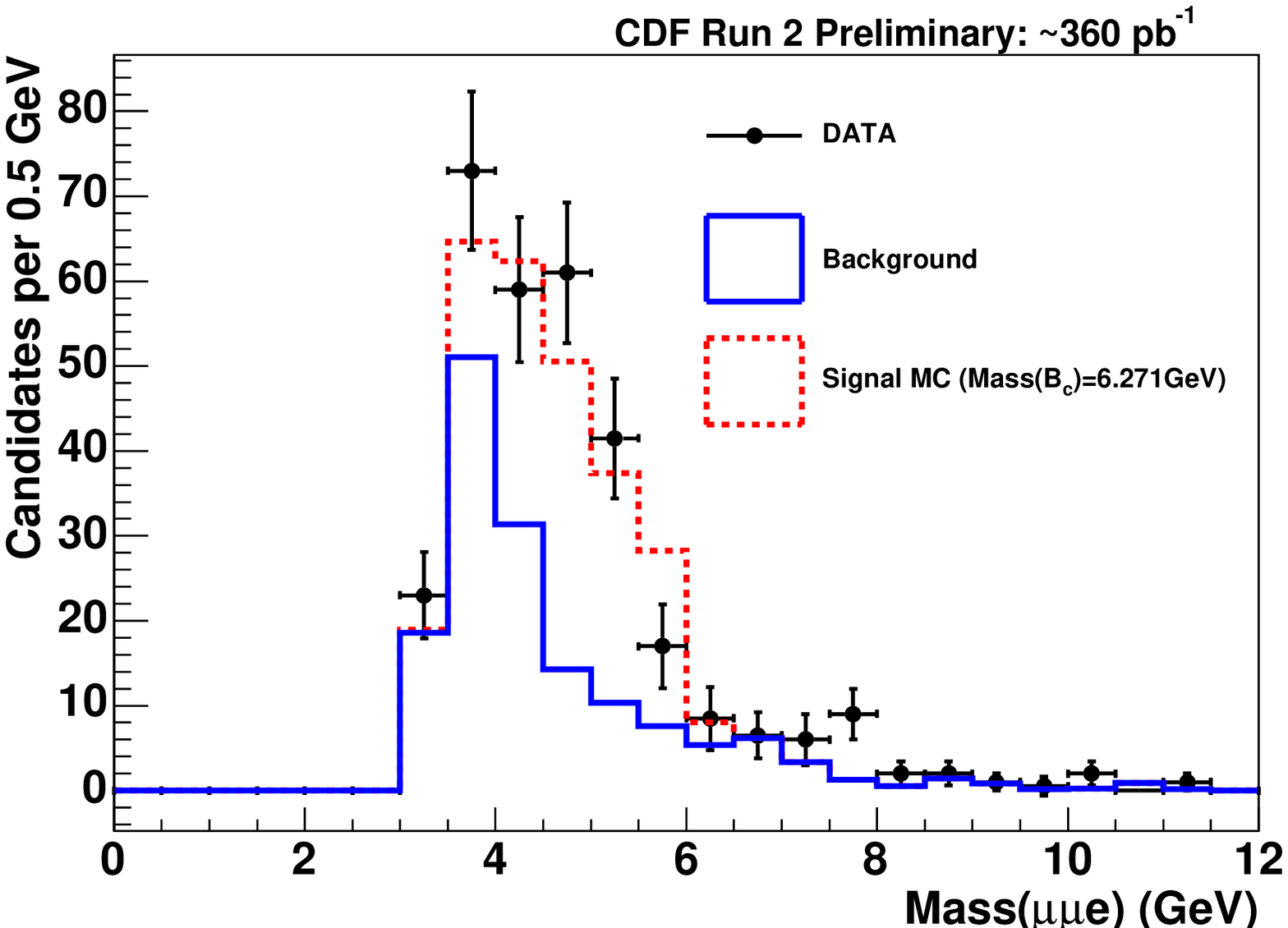}
 \hspace{-0.8cm}
 \includegraphics[height=.28\textheight]{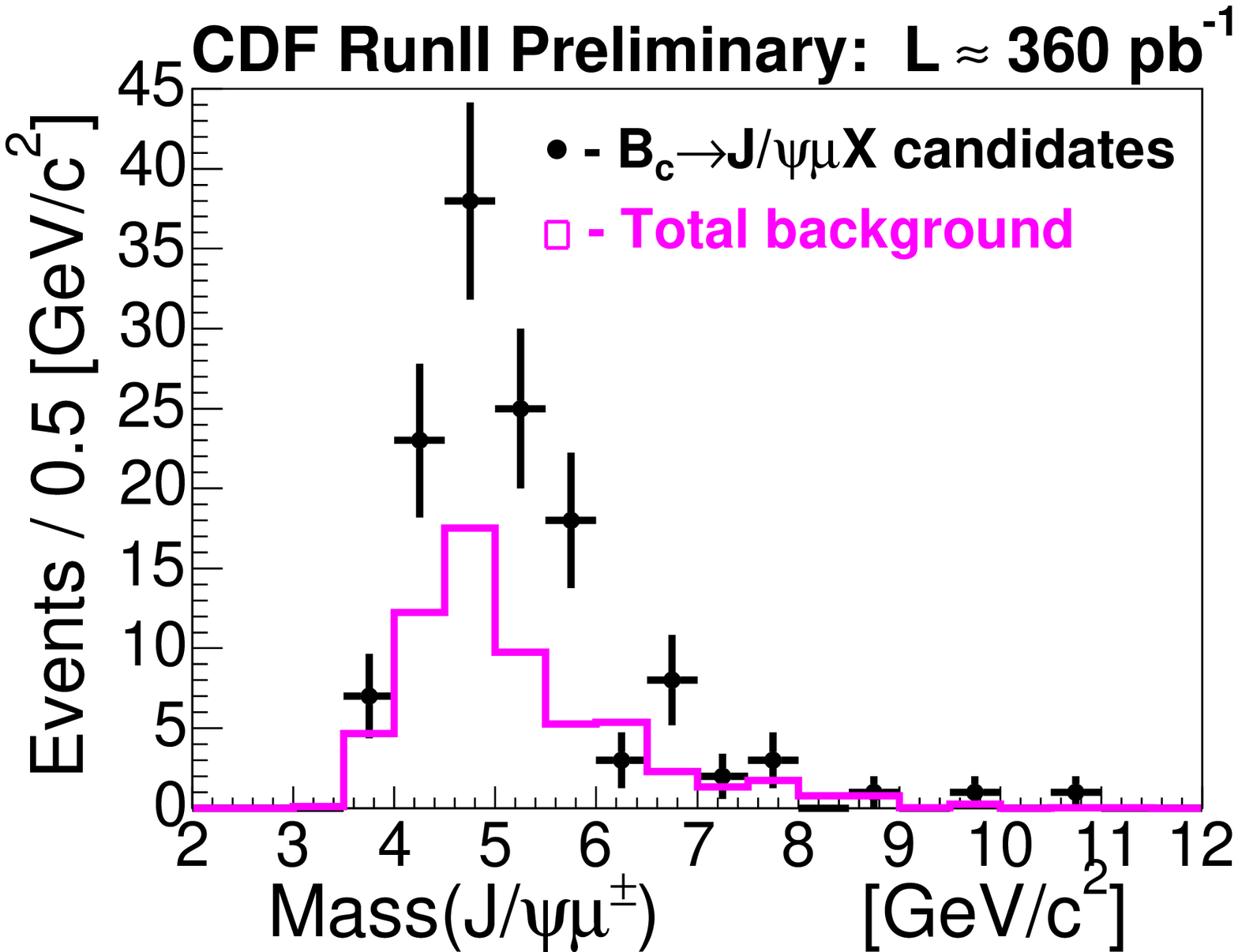}
  \caption{(left) $J/\psi e$ mass distribution for total background and 
data.  (right) $J/\psi \mu$ mass distribution for total background and data.}
\end{figure}

Using 360 pb$^{-1}$ of Run II data CDF performed the $B_c^{\pm}$ search in 
semileptonic channels as well. A signal of 
114.9 $\pm$ 15.5 (stat) $\pm$ 13.6 (syst) events (5.9 $\sigma$) was
observed in the $B_c^{\pm} \rightarrow J/\psi e^{\pm} \nu$ 
channel (see Fig. 8(left)).
The cross section times the branching fraction for 
$B_c^{\pm} \rightarrow J/\psi e^{\pm} \nu$ relative to 
$B^{\pm} \rightarrow J/\psi K^{\pm}$ was measured to be 
0.282 $\pm$ 0.038(stat) $\pm$ 0.074 (syst) in the kinematic region
of $p_T(B)>$ 4.0 GeV/c and $|y(B)| <$ 1.0.
Similarly, an excess of 
59.1 $\pm$ 12.5 events (5.2 $\sigma$) was observed in the 
$B_c^{\pm} \rightarrow J/\psi \mu^{\pm} \nu$ (see Fig. 8(right)).
The cross section times the branching fraction for 
$B_c^{\pm} \rightarrow J/\psi \mu^{\pm} \nu$ relative to 
$B^{\pm} \rightarrow J/\psi K^{\pm}$ was measured to be 
0.249 $\pm$ 0.045(stat) $^{+0.107}_{-0.076}$ (syst) in the kinematic region
of $p_T(B)>$ 4.0 GeV/c and $|y(B)| <$ 1.0.

\section{4. $b$-baryons}

As discussed earlier, using 220 pb$^{-1}$ of Run II data CDF was able to 
present the most precise measurement, to date, for the
$\Lambda^0_b$ mass \cite{CDFbmasses}. 
The $\Lambda_b$ mass measurement was made by reconstructing 89$\pm$10 signal 
events in the decay channel
$\Lambda_b^0 \rightarrow J/\psi \Lambda^0$ where $\Lambda^0 \rightarrow p\pi^-$ 
and 
$J/\psi \rightarrow \mu^+\mu^-$.
The mass
was measured to be $m(\Lambda_b) =$ 5619.7 $\pm$ 1.2(stat) $\pm$ 1.2(syst) 
MeV/$c^2$. This measurement has better uncertainty than the world average.

\begin{figure}
  \includegraphics[height=.3\textheight]{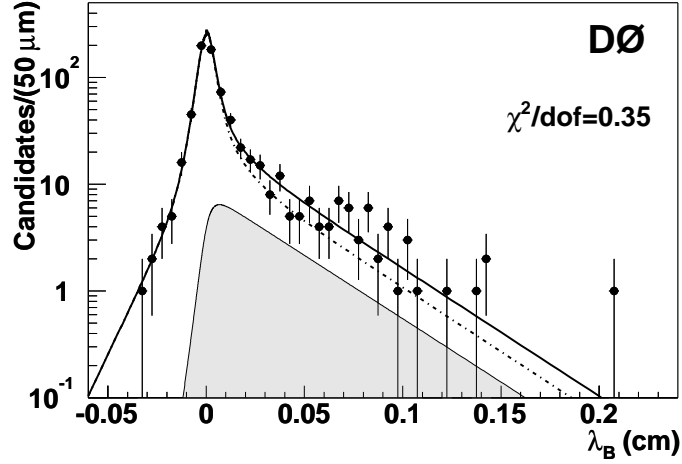}
  \caption{Proper decay length distribution for $\Lambda_b^0$ candidates. The
points are the data, and the solid curve is the sum of fitted contributions 
from signal (gray) and the background (dashed-dotted line).}
\end{figure}

Using 250 pb$^{-1}$ of Run II data D0 reconstructed 61$\pm$12 signal events 
in the exclusive decay channel $\Lambda_b^0 \rightarrow J/\psi \Lambda^0$ 
where $\Lambda^0 \rightarrow p\pi^-$ and 
$J/\psi \rightarrow \mu^+\mu^-$ and 
measured the $\Lambda_b^0$ lifetime. 
 The $\Lambda_b^0$ lifetime was determined to 
be 1.22$^{+0.22}_{-0.18}$ (stat)$\pm$0.04(syst) $ps$ \cite{D0_Lambdab} 
(see Fig. 9).
The winter 2005 HFAG world average for $\tau (\Lambda_b^0)$ was 
1.232$\pm$0.072 $ps$ and $\tau(\Lambda_b^0)/\tau(B^0)$ equal to
0.806$\pm$0.047 \cite{HFAG_2005}.
The $\Xi_b$ lifetime has not been updated recently. The winter 2005 HFAG 
world average is equal to 1.39$^{+0.34}_{-0.28}$ $ps$ and it refers to a 
mixture of $\Xi_b^0$ and $\Xi_b^{\pm}$. This value is based on older 
measurements performed by the DELPHI and ALEPH experiments.

\begin{figure}
  \includegraphics[height=.28\textheight]{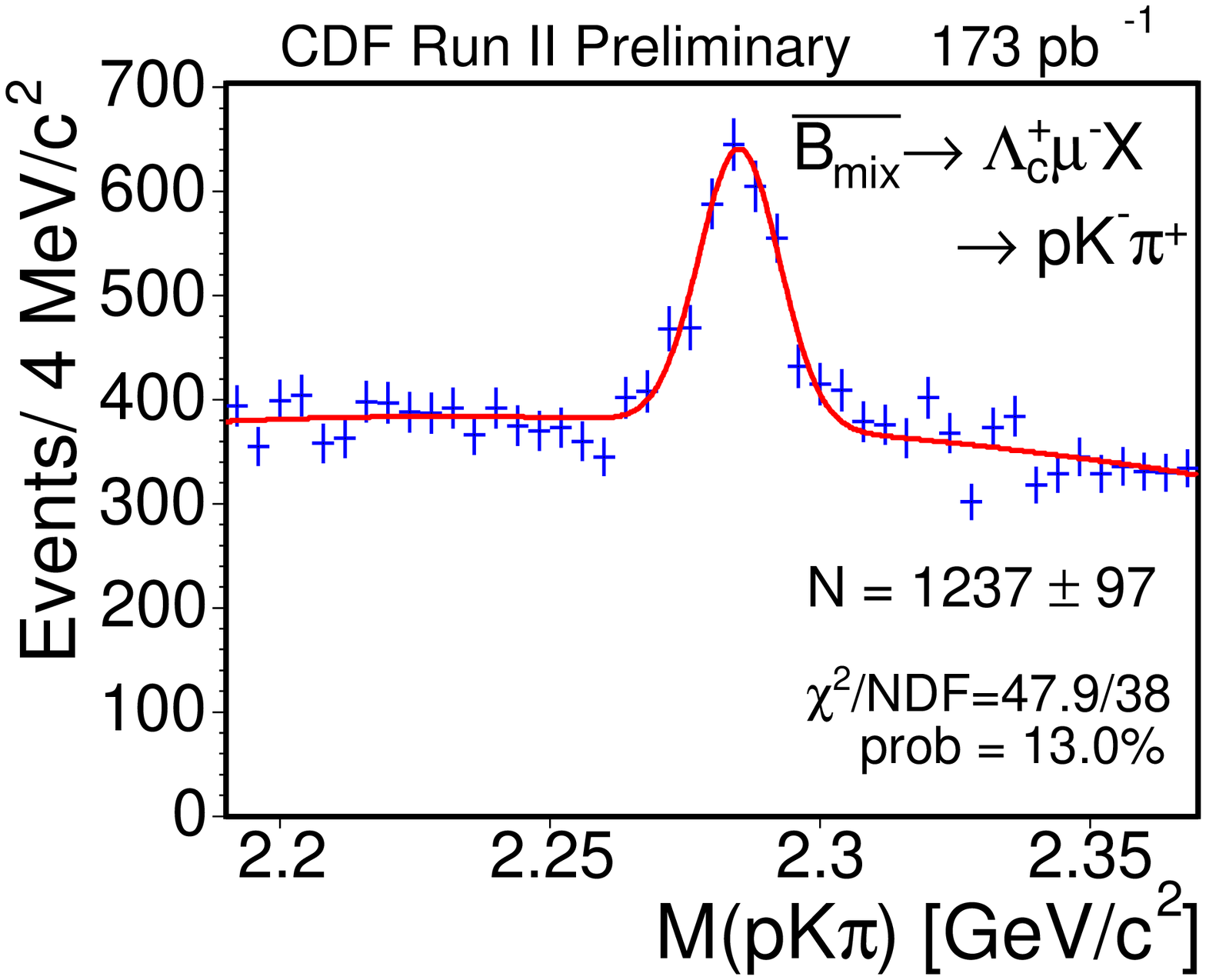}
 \hspace{-0.1cm}
 \includegraphics[height=.28\textheight]{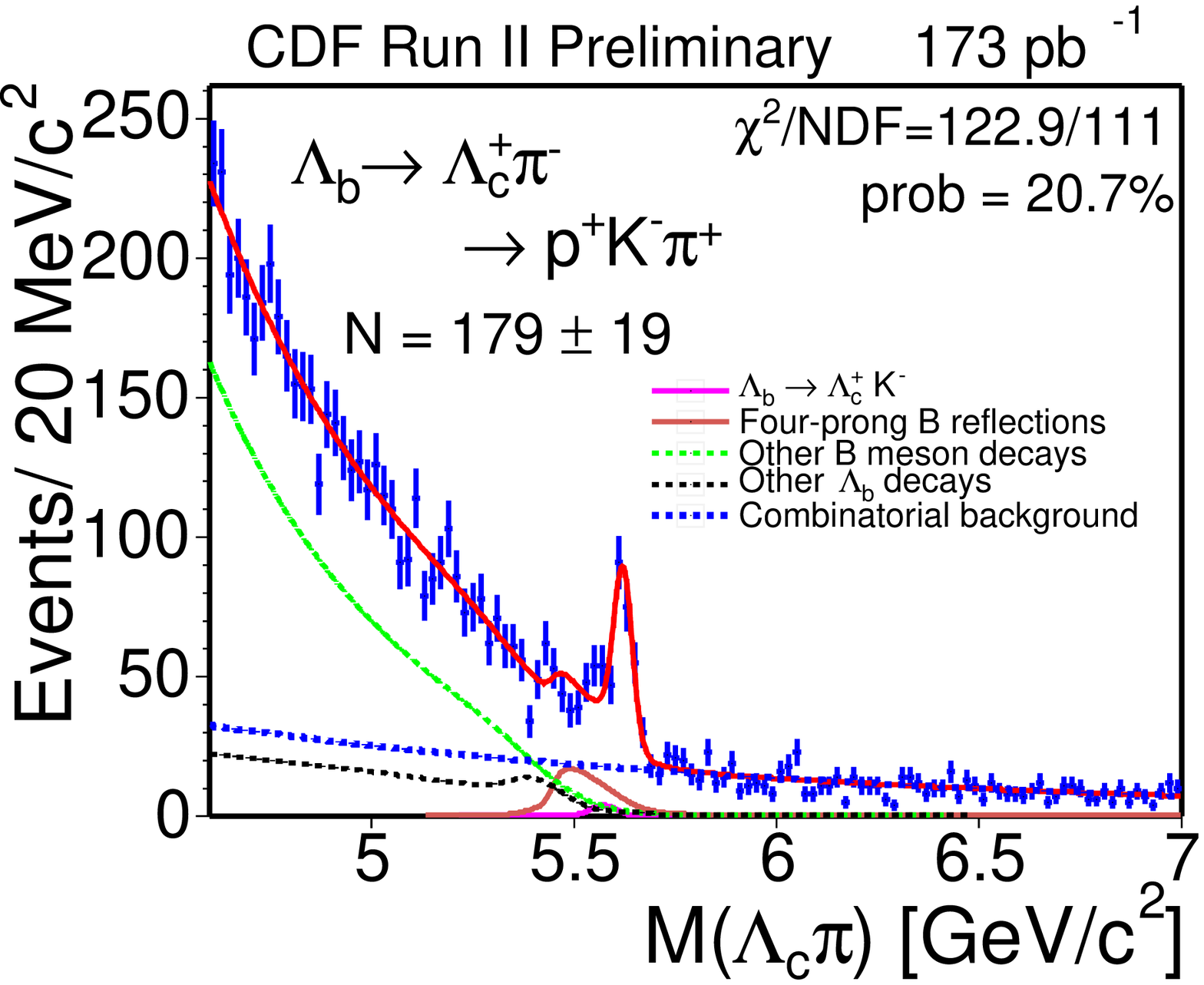}
  \caption{(left) The M(pK$\pi$) invariant mass distribution (right) 
The M($\Lambda_c \pi$) invariant mass distribution. The results of an unbinned
likelihood fit are superimposed on the histograms and a $\chi^2$ probability
is calculated.}
\end{figure}

Using a Run II sample of 173 pb$^{-1}$ CDF presented the first measurement
of the ratio of branching fractions
$Br(\Lambda_b \rightarrow \Lambda_c^+ \mu^-\bar{\nu_{\mu}})$/
$Br(\Lambda_b \rightarrow \Lambda_c^+ \pi^-)$ which is a good 
test of HQET. This analysis used the displaced vertex trigger and 
reconstructed 1237$\pm$97 $\bar{B} \rightarrow \Lambda_c^+\mu^- X$ decays 
(see Fig. 10(left)) and 179$\pm$19 
$\Lambda_b \rightarrow \Lambda_c^+\pi^-$ decays (see Fig. 10(right)).
In the process of this analysis CDF also observed several $\Lambda_b^0$ 
semileptonic decays which have never been seen before: 
$\Lambda_b^0 \rightarrow \Lambda_c(2593)^+\mu^- X$, 
$\Lambda_b^0 \rightarrow \Lambda_c(2625)^+\mu^- X$,
$\Lambda_b^0 \rightarrow \Sigma_c^0\pi^+\mu^- X$, and
$\Lambda_b^0 \rightarrow \Sigma_c^{++}\pi^-\mu^- X$.
They measured the
$Br(\Lambda_b^0 \rightarrow \Lambda_c^+ \mu^-\bar{\nu_{\mu}})$/
$Br(\Lambda_b^0 \rightarrow \Lambda_c^+ \pi^-)$  
to be equal to
20.0$\pm$3.0(stat)$\pm$1.2 (syst)$^{+0.7}_{-2.1}$(BR)$\pm$0.5 (UBR), where
UBR stands for unmeasured branching fractions.

\section{5. Conclusions-Prospects}

More than 1 fb$^{-1}$ of $p\bar{p}$ collisions has been already 
delivered by the Tevatron,
and 
many new results have been recently produced by the CDF and D0 experiments.
As the Tevatron is expected to provide between 4.1 and 8.2 
fb$^{-1}$ by October 2009, a lot of answers to explored and yet unexplored 
questions and a lot of surprises are awaiting.  


\begin{theacknowledgments}
  I would like to thank the conference organizers for a very productive 
meeting. I would also like to thank my colleagues  
M. Herndon, D. MacFarlane, S. Nahn and R. Van Kooten 
for discussions on the results presented here.
\end{theacknowledgments}




\bibliography{sample}




\end{document}